\documentclass[
aip,
rsi,
amsmath,amssymb,
reprint,
]{revtex4-1}

\usepackage{graphicx}
\usepackage{dcolumn}
\usepackage{bm}
\usepackage{float}
\usepackage[caption = false]{subfig}
\usepackage{natmove}
\usepackage[title]{appendix}

\newcommand{\tr}{\text{Tr}}

\newcommand{\ket}[1]{| #1 \rangle}
\newcommand{\bra}[1]{\langle #1 |}

\begin{document}

\title{Simulating conical intersection dynamics in the condensed phase with hybrid quantum master equations}

\author{Addison J. Schile}
\affiliation{Department of Chemistry, University of California, Berkeley}
\affiliation{Lawrence Berkeley National Laboratory, University of California, Berkeley}
\author{David T. Limmer}
 \email{dlimmer@berkeley.edu.}
\affiliation{Department of Chemistry, University of California, Berkeley}
\affiliation{Kavli Energy NanoSciences Institute, University of California, Berkeley}
\affiliation{Lawrence Berkeley National Laboratory, University of California, Berkeley}

\date{\today}

\begin{abstract}
We present a framework for simulating relaxation dynamics through a conical intersection of an open quantum system that combines methods to approximate the motion of degrees of freedom with disparate time and energy scales. In the vicinity of a conical intersection, a few degrees of freedom render the nuclear dynamics nonadiabatic with respect to the electronic degrees of freedom. We treat these strongly coupled modes by evolving their wavepacket dynamics in the absence of additional coupling exactly. The remaining weakly coupled nuclear degrees of freedom are partitioned into modes that are fast relative to the nonadiabatic coupling and those that are slow. The fast degrees of freedom can be traced out and treated with second-order perturbation theory in the form of the time-convolutionless master equation. The slow degrees of freedom are assumed to be frozen over the ultrafast relaxation, and treated as sources of static disorder. In this way, we adopt the recently developed frozen-mode extension to second-order quantum master equations. We benchmark this approach to numerically exact results in models of pyrazine internal conversion and rhodopsin photoisomerization. We use this framework to study the dependence of the quantum yield on the reorganization energy and the characteristic timescale of the bath, in a two-mode model of photoisomerization. We find that the yield is monotonically increasing with reorganization energy for a Markovian bath, but monotonically decreasing with reorganization energy for a non-Markovian bath. This reflects the subtle interplay between dissipation and decoherence in conical intersection dynamics in the condensed phase.
\end{abstract}

\maketitle

\section{\label{sec:intro}Introduction}

The ultrafast excited-state relaxation dynamics of polyatomic systems are nearly universally mediated by motion through conical intersections\cite{worth2004beyond,schuurman2018dynamics}. Advanced time-dependent spectroscopies have made the identification of such molecular motions possible in principle\cite{chachisvilis1999femtosecond,neumark2008slow,oliver2014correlating,kowalewski2015catching,domcke2012role}, elucidating their role in many photochemical reactions\cite{kitney2014two,devine2016non,wu2019two}. Theory and simulation are useful tools to interpret and elucidate the microscopic motions associated with the degrees of freedom probable with experiment. However, the ability to accurately and efficiently simulate such nonadiabatic dynamics in the condensed phase is challenging\cite{hughes2009effective1,hughes2009effective2,tully2012perspective}. Nonadiabatic systems by definition contain many strongly coupled nuclear and electronic degrees of freedom, blurring the separation of time scales between their motion, and demanding a quantum description of both. In the condensed phase, the ability to correctly describe dissipation requires that a bath is represented either implicitly or explicitly, complicating approximations that make such calculations tractable in the gas phase. Here, we describe a framework to leverage a separation of energy and time scales to arrive at a hybrid method to study the dynamics of molecules through conical intersections. The method we employ treats the most strongly coupled modes explicitly, and develops a hybrid reduced description for the remaining modes by identifying some as slow and others as fast relative to the nonadiabatic dynamics. This methodology enables us to study photochemical quantum yields in widely different environments. 

Existing approaches to study motion through conical intersections in condensed phases fall into two broad categories. One way is to represent all of the interacting degrees of freedom and compute the dynamics of a closed system, albiet one with a large enough number of states to approximate the environment. These sorts of approaches can range from numerically exact methodologies such as the Multiconfigurational Time-Dependent Hartree (MCTDH) method\cite{beck2000multiconfiguration,wang2003multilayer}, the Quasiadiabatic Path Integral method\cite{topaler1993quasi,topaler1996path}, and multiple spawning techniques\cite{ben1998nonadiabatic,levine2007isomerization} to more approximate methods such as mixed-quantum classical theories like Ehrenfest\cite{kapral2006progress} and surface hopping\cite{tully1990molecular,landry2012recover,kelly2013efficient}, and semiclassical theories as obtained from the mapping approaches.\cite{meyer1979classical,stock1997semiclassical,thoss1999mapping} Exactly representing the degrees of freedom has the advantage that arbitrary degrees of freedom can be represented provided a large enough basis. While these tools have shown promise in a variety of systems, they still can be limited by severe exponential scaling in numerically exact approaches or by invoking uncontrolled approximations that can break fundamental symmetries like detailed balance, complicating the description of a thermalizing bath. 

An alternate approach relies on the master equation approach of open quantum systems\cite{nakajima1958quantum,zwanzig1960ensemble,breuer2002theory}, in which the dynamics of only a few relevant degrees of freedom are represented explicitly in a reduced density matrix that is coupled implicitly a set of environmental degrees of freedom. Often the environment is taken as an infinite bath of harmonic degrees of freedom, though this is not required. When a harmonic bath is used, these methods assume that a linear response relationship between the system and bath holds and thus the bath represents a set of degrees of freedom that obey gaussian statistics. When this approximation is valid, these methods also have a range of accuracy from the numerically exact Hierarchical Equation of Motion (HEOM)\cite{tanimura1989time} to perturbative treatments such as Redfield theory.\cite{redfield1957theory} By construction, most of these approaches accurately describe dissipation to the environment. However, they suffer from pitfalls in computational complexity, as HEOM scales roughly factorially in the system-bath coupling strength, or accuracy, as many perturbative theories have known issues with preserving the trace and positivity of the reduced density matrix.

In this paper, we propose the use of a hybrid methodology, in the spirit of previous work\cite{thoss2001self,berkelbach2012reduced1,berkelbach2012reduced2}, in which both approaches are utilized in regimes where they are valid. The most strongly coupled, anharmonic degrees of freedom are evolved directly and the remaining degrees of freedom are treated with different approximate theories, whose applicability rests in identifying relevant separations of time and energy scales. This approach has the advantage of a reduced computational cost compared to the most demanding numerically exact methods, while retaining both flexibility and accuracy and relies heavily on recent work in applying the so-called frozen mode approximation to quantum master equations.\cite{montoya2015extending} The present paper is organized in four remaining sections. In Sec. \ref{sec:theory}, the general framework for developing a hybrid method in the context of conical intersection models is outlined. In Sec. \ref{sec:results}, this methodology is benchmarked in models of internal conversion of pyrazine and photoisomerization of rhodopsin by comparing to existing numerically exact results.  In Sec. \ref{Sec:Yields} we apply the framework to address the dependence of the quantum yield on the environment. Some concluding remarks are given in Sec. \ref{sec:conclusions}.

\section{\label{sec:theory}Theory}

In this section we describe the framework on which a hybrid methodology can be built. This framework can begin from an \textit{ab initio} molecular Hamiltonian, provided a diabatic basis can be constructed that minimizes the nonadiabatic coupling from the kinetic energy derivatives.\cite{kouppel1984multimode,van2010diabatic} In the diabatic basis we can write the Hamiltonian as,
\begin{equation}
H = \sum_{i,j} \ket{i} \left[ T(\mathbf{Q}) \delta_{ij} + V_{ij} (\mathbf{Q}) \right] \bra{j},
\end{equation}
where $T(\mathbf{Q})$ is the kinetic energy operator, which is diagonal, $V_{ii}(\mathbf{Q})$ is the potential energy surface of the $i$th diabatic electronic state and $V_{i\ne j}(\mathbf{Q})$ is the diabatic coupling between states $i$ and $j$ with $\mathbf{Q} = \{ Q_1, Q_2, \ldots, Q_N \}$ the vector of displacements of each $N$ nuclear degree of freedom from a reference geometry, $\mathbf{Q}_0$, or generalized modes. In principle, the full system can be completely described at all times by its density matrix, 
$\rho (t)$, whose time evolution is given by the Liouville-von Neumann equation
\begin{equation}
\partial_t \rho (t) = -i [ H, \rho (t) ]
\end{equation}
where $[\cdot, \cdot]$ is the commutator.
Due to exponential scaling of standard basis set treatments, this description becomes intractable for systems beyond only a few degrees of freedom, and in the condensed phase reduced descriptions are required. Throughout we will set $\hbar=1$ and use mass weighted coordinates unless otherwise explicitly stated.

\subsection{Mode expansion}
To build a reduced description of the dynamics, we first impose some structure on the many body potential $V_{ij} (\mathbf{Q})$ appropriate for a molecule in a surrounding environment with a conical intersection. Within a general mode expansion\cite{kouppel1984multimode}, $V_{ij} (\mathbf{Q})$ can be approximated as,
\begin{align}
V_{ij} (\mathbf{Q}) = &V^{(0)}_{ij} + \sum_k  V^{(1)}_{ij}(Q_k) + \sum_{k < l}  V^{(2)}_{ij}(Q_k,Q_l) + \dots
\end{align}
where $V^{(n)}_{ij}$ is a potential function that couples $n$ modes of the system, truncated here to second order. Generally, each order potential could be a distinct function of its arguments, whose repeated indices we suppress for clarity. 

As we are interested in motion in the vicinity of a conical intersection, we will isolate two orthogonal coordinates, the tuning mode, $q_t$, and a coupling mode, $q_c$, which define a surface of points where the two potential energy surfaces $i$ and $j$ intersect, giving rise to large non-adiabatic coupling. In the following, these are the modes we will consider strongly coupled. In principle, additional modes with coupling constants large relative to the bare electronic energy gap, or modes with large amplitude motion, should be included in this description. For the models we study only these two coordinates are included. 

We will assume that only the tuning mode undergoes large amplitude motion away from the reference geometry. Under such assumption, which could be relaxed, we have a potential for the tuning mode of the form,
\begin{equation}
V^{(1)}_{ij}(q_t) = \delta_{ij} \left (v_i(q_t) + \kappa_t^{(i)} q_t  \right )
\end{equation}
where $v_i(q_t)$ is in general anharmonic. We assume the coupling mode is harmonic, 
\begin{equation}
V^{(1)}_{ij}(q_c) = \delta_{ij} \left (\frac{1}{2} \Omega_c q_c^2 + \kappa_c^{(i)} q_c  \right ) + (1-\delta_{ij}) \lambda^{(ij)} q_c 
\end{equation}
with frequency, $\Omega_c$, is given by,
\begin{equation}
\Omega_c = \left( \frac{\partial^2 V_{ii}}{\partial q_c^2} \right)_{\mathbf{Q}_0},
\end{equation}
where the constants $V^{(0)}_{ij}$ are defined by the reference geometry $\mathbf{Q}_0$. We pull out the linear portions of the potentials, parameterized by $\kappa_k^{(i)}$, for clarity, which are Holstein-like coupling coefficients given by,
\begin{equation}
\kappa_k^{(i)} =  \left( \frac{\partial V_{ii}(\mathbf{Q})}{\partial q_k} \right)_{\mathbf{Q}_0},
\end{equation}
and $\lambda^{(ij)}$ is a Peierls-like coupling coefficient, given by
\begin{equation}
\lambda^{(ij)} =  \left( \frac{\partial V_{ij}}{\partial q_c} \right)_{\mathbf{Q}_0}
\end{equation}
which is the only off-diagonal term in the diabatic state basis we consider and due to hermiticity, $\lambda^{(ij)} = \lambda^{(ji)}$. The existence of both $\lambda^{(ij)}$ and the $\kappa_k^{(i)}$'s reflect that at a conical intersection, both the electronic gap as well as the electronic coupling are modulated by nuclear degrees of freedom. The remaining modes are assumed to be harmonic,
\begin{equation}
V^{(1)}_{ij}(Q_k) =  \frac{1}{2} \omega_k Q_k^2 + c_{k}^{(i)} Q_k
\end{equation}
with frequencies,
\begin{equation}
\omega_k = \left( \frac{\partial^2 V_{ii}}{\partial Q_k^2} \right)_{\mathbf{Q}_0} \, ,
\end{equation}
and additional Holstein couplings,
\begin{equation}
c_{0,k}^{(i)} =  \left( \frac{\partial V_{ii}(\mathbf{Q})}{\partial Q_k} \right)_{\mathbf{Q}_0},
\end{equation}
for each $i$th electronic state.

Provided the linear response form for all of the modes not including $q_t$, the highest-order mode coupling potential we consider that is consistent with this choice is bilinear in the modes. Specifically, we take
\begin{equation}
V^{(2)}_{ij}(Q_k,Q_l) = c_{k,l}^{(i)} Q_k Q_l \delta_{ij}(1-\delta_{lk})
\end{equation}
where $c_{k,l}^{(i)}$ is the coupling coefficient that transfers vibrational energy between the $k$th and $l$th modes,
\begin{equation}
c_{k,l}^{(i)} = \left( \frac{\partial^2 V_{ii}}{\partial Q_k \partial Q_l} \right)_{\mathbf{Q}_0} \, ,
\end{equation}
which we take as diagonal in the diabatic states. By construction this is zero between the tuning and coupling modes, as these are chosen to be orthogonal. With the exception of the tuning mode, the remaining coordinates are all harmonic, so we can in principle orthogonalize the remaining $N-2$ subspace defined outside of $q_t$ and $q_c$. The enables us to set $c_{k,l}^{(i)}$ to zero for all $l$ and $k$ that do not include $q_c$ or $q_t$. 

The resultant potential has a simple approximate form. The diabatic coupling is given by
\begin{equation}
V_{i\ne j} (\mathbf{Q}) = \lambda^{(ij)} q_c
\end{equation}
containing only the coupling mode with Peierls constant, where here we have taken $V^{(0)}_{i \ne j}=0$. The diabatic potentials are given by
\begin{align}
V_{ii} (\mathbf{Q}) &=  V^{(0)}_{ii} + v_i(q_t) + \kappa_t^{(i)} q_t   + \frac{1}{2} \Omega_c q_c^2 + \kappa_c^{(i)} q_c \nonumber \\
&+ q_t \sum_{k} c_{t,k}^{(i)} Q_k + q_c \sum_{k} c_{c,k}^{(i)} Q_k + \sum_{k} c_{0,k}^{(i)} Q_k   \nonumber \\
&+  \sum_{k} \frac{1}{2} \omega_k Q_k^2 
\end{align}
where the tuning and coupling coordinates can exchange energy with the remaining $N-2$ modes, in such a way as to renormalize the effective Holstein and Peierls couplings. This potential is envisioned as including only the minimal ingredients required to describe a conical intersection with a surrounding environment, as additional complexity could be added if any of the approximations above were found invalid.\cite{kouppel1984multimode}

\subsection{System bath partitioning}

For an isolated system, the $N$ mode diabatic potential described above can be simulated directly using compact basis set techniques like MCTDH and multiple spawning. \cite{beck2000multiconfiguration,wang2003multilayer,ben1998nonadiabatic,levine2007isomerization}
 However, in a condensed phase, in order to correctly describe dissipation and relaxation, we require that the number of modes goes to infinity, such that the $\omega_k$'s will form a continuous band of frequencies. While basis set techniques can approximate this continuous band, doing so typically results in algorithms that scale exponentially in time\cite{bonfanti2012compact}.
As the remaining environment modes are expected to be less strongly coupled to the electronic degrees of freedom, we can consider ways to integrate them out and arrive at a reduced description of the dynamics of the system. In this way, we will define the total Hamiltonian, $H=H_S+H_{SB}+H_B$, as a partitioning between a system, bath and coupling terms. To determine an effective partitioning, we can leverage the identification of the relevant coupling constants and their expected scales.

Since we expect the coupling and tuning modes to be strongly coupled to the electronic states, we will treat their dynamics in the absence of additional coupling exactly. Restricting ourselves to two diabatic states, we refer to them, along with the electronic states, as the system Hamiltonian, $H_S$,
\begin{align}
H_S &= \sum_{i,j=1,2} \ket{i} h_i \delta_{ij}  + \lambda q_c (1-\delta_{ij}) \bra{j} \nonumber \\
h_i &= T+V^{(0)}_{ii}+v_i(q_t) + \kappa_t^{(i)} q_t   + \frac{1}{2} \Omega_c q_c^2 + \kappa_c^{(i)} q_c
\end{align}
where $T$ is the kinetic energy of the tuning and coupling modes and we have removed the electronic state dependence from $\lambda$. In the case that both coordinates are harmonic, this Hamiltonian reduces to the so-called linear vibronic model\cite{kouppel1984multimode}. 
The remaining degrees of freedom, the $Q_k$'s, will make up a bath portion of the Hamiltonian. The coupling between the bath degrees of freedom and the system will be denoted by the system-bath coupling Hamiltonian, $H_{SB}$. This term can be written in the direct product form,
\begin{align}
H_{SB} &= \sum_{n=0,c,t} s_n \sum_{k} c_{n,k}^{i} Q_k \\
s_{(c,t)} &= \sum_i  \ket{i}q_{(c,t)} \bra{i} \, , \quad \quad s_0 = \sum_i  \ket{i} \bra{i} \nonumber \, ,
\end{align}
where $s_n$ in general includes both direct coupling to the electronic system and vibrational relaxation through coupling to the tuning or coupling modes.
To describe the system-bath coupling strengths, it is useful to define the spectral densities, $J_n (\omega)$, for each system bath operator,  
\begin{equation}
J_n (\omega) = \frac{\pi}{2} \sum_k \frac{c^2_{n,k}}{\omega_k} \delta (\omega - \omega_k).
\end{equation}
and are parameterized by a reorganization energy, $E_{r,n} $,
\begin{equation}
E_{r,n} = \frac{1}{\pi} \int_0^{\infty} d \omega \; \frac{J_n(\omega)}{\omega},
\end{equation}
and a characteristic frequency, $\omega_{c,n}$. The reorganization energy reflects the overall strength of the coupling of the system to the bath, and the characteristic frequency determines the decay of the spectral density at infinite frequency. In order to treat the bath perturbatively, the dimensionless coupling parameter, $\eta$, given by
\begin{equation}
\eta =\max_n \left [ \frac{2}{\pi^2 \omega_{c,n}} \int_0^\infty d\omega \frac{J_n(\omega)}{\omega} \right ]
\end{equation}
must be small on an absolute scale, $\eta \ll 1$. This parameter reflects the competing effects of the reorganization energy and characteristic frequency on the decay of higher-order correlation functions used in a perturbative expansion, and can be derived explicitly for simple models.\cite{montoya2015extending,laird1991quantum} For fixed $E_{r,n}$, $\eta$ increases as $\omega_{c,n}$ gets smaller, generally violating the criteria for perturbation theory. This scaling of $\eta$ with $\omega_{c,n}$ makes it difficult to use standard quantum master equation approaches for studying motion through conical intersections, as the relevant scale of the system dynamics is ultrafast, rendering typical bath relaxation times comparatively long\cite{gindensperger2006short1,gindensperger2006short2}. To remedy this requires confronting non-Markovian effects directly.

The remaining terms in the Hamiltonian are labeled as the bath, $H_B$, and are given by a set of noninteracting harmonic oscillators,
\begin{align}
H_B &= \frac{1}{2} \sum_{k\in \mathrm{slow}} \omega_{k} \left( -\frac{\partial^2}{\partial Q_k^2} +  Q_{k}^2 \right) \nonumber \\
& + \frac{1}{2} \sum_{k\in \mathrm{fast}} \omega_{k} \left( -\frac{\partial^2}{\partial Q_k^2} +  Q_{k}^2 \right) \, ,
\end{align}
which we will partition into a group labeled \emph{fast} and a group labeled \emph{slow}, depending on the oscillator's frequency, $\omega_k$ relative to a parameter $\omega^*$.  Here $\omega^*$ is a frequency that delineates between the fast and slow modes of the bath relative to a characteristic time scale of the system. As motion through a conical intersection is mediated by the nonadiabatic coupling, we assume the characteristic time scale of the system to be given by the Peierls coupling, $\lambda$, and consider slow modes to be those with $\omega_k < \lambda$. 

\subsection{Hybrid dynamical approach}

Given the system-bath partitioning proposed above, we can develop an approximate way to evolve a reduced system dynamics that is capable of correctly describing dissipation even when some of the bath degrees of freedom are non-Markovian owing to the large separation of timescales between non-adiabatic system dynamics and slow environmental motions. To this aim we follow the procedure outlined in Ref. \onlinecite{montoya2015extending}. Specifically, we consider the time dependent reduced density matrix, $\sigma(t)$, as
\begin{equation}
\sigma(t) = \mathrm{Tr}_B \{ \rho(t) \}
\end{equation}
where the trace is taken over all $\mathbf{Q}$ defined in the bath part of the Hamiltonian. In order to obtain a closed evolution equation for $\sigma(t)$, we leverage the expected separation of timescales between evolution in the fast part of the bath and those in the slow part of the bath.

Following the partitioning in $H_B$, we can similarly partition a given spectral density into the slow and fast portions,\cite{berkelbach2012reduced1,berkelbach2012reduced2,montoya2015extending}
\begin{equation}
J_n (\omega) = J_{n,\text{slow}} (\omega) + J_{n,\text{fast}} (\omega).
\end{equation}
where 
\begin{equation}
J_{n,\text{slow}} (\omega) = S(\omega)J_n (\omega)
\end{equation}
delineates the slow portion and 
\begin{equation}
J_{n,\text{fast}} (\omega) = [1-S(\omega)]J_n (\omega)
\end{equation}
the fast portion where
\begin{equation}
S(\omega)= 
\begin{cases} (1-(\omega/\omega^*)^2)^2 \quad \omega < \omega^*\\
0 \quad \omega \ge \omega^*
\end{cases}
\end{equation}
is a splitting function, parameterized by $\omega^*$. 
In the limit that $\omega^* \ll \lambda$, we can consider the slow modes as static over the course of system dynamics. In such a limit, the total time-dependent density matrix factorizes into an initial piece from the slow modes and a time dependent remainder in which the fast modes of the bath and the system degrees of freedom evolve, $\rho(t) \approx \rho_{\boldsymbol{Q} \in \mathrm{slow}}(0) \rho_{\boldsymbol{Q} \in \mathrm{fast},S}(t)$.  In such a case, the slow modes contribute only as a source of quenched disorder to the system Hamiltonian, and induces an inhomogeneous broadening due to different realizations of initial conditions. We include this part of the system-bath coupling directly into the Hamiltonian as,
\begin{equation}
\tilde{H}_S = H_S + \sum_n s_n \sum_{k \in \text{slow}} c_{n,k} \tilde{Q}_{k},
\end{equation}
where $\tilde{Q}_{k}$ is a classical variable, not an operator. Since these modes are incorporated into the system Hamiltonian, provided the assumed separation of time scales holds, they are treated to all orders in their coupling strength. 

The reduced density matrix is obtained by averaging over different realizations of the reduced density matrix corresponding to different realizations of initial conditions,
\begin{equation}
\sigma (t) = \int d \mathbf{\tilde{Q}} \, p ( \mathbf{\tilde{Q}} )  \tilde{\sigma} (t),
\end{equation}
 where 
 \begin{equation}
\tilde{\sigma} (t) = \mathrm{Tr}_{\mathbf{{Q}} \in \mathrm{fast}} \{ \rho(t) \}
\end{equation}
is the reduced density matrix computed by tracing over only the fast degrees of freedom, which depends parametrically on the slow bath degrees of freedom. The initial conditions of the slow modes are drawn from the distribution $p ( \mathbf{\tilde{Q}} )$. Depending on the temperature relative to the characteristic frequency of the slow bath modes, $p ( \mathbf{\tilde{Q}} )$ may be a Wigner distribution or a Boltzmann distribution. 
 
The remaining modes in the bath now have a smaller overall reorganization energy. Since by construction, these modes relax on a timescale faster than the system, they induce Markovian or nearly-Markovian dissipation and decoherence. For this reason, we can treat these degrees of freedom with time-dependent Redfield theory, also known as the 2nd-order Time Convolutionless master equation\cite{shibata1977generalized,chaturvedi1979time} (TCL2). In the TCL2 formalism, the dynamics of the each realization of the reduced density matrix are given by,\cite{pollard1994solution}
\begin{align}
\partial_t \tilde{\sigma} (t) &= -i [\tilde{H}_S, \tilde{\sigma} (t)] \;+ \nonumber \\ & \sum_n [\Theta_n (t) \tilde{\sigma} (t) , s_n] + [s_n, \tilde{\sigma} (t) \Theta^{\dagger}_n (t)],
\end{align}
where $\Theta_n (t)$ is the system operator dressed by the time-dependent rates given by the bath correlation function. In the eigenstate basis of $\tilde{H}_S$, each element is given by
\begin{equation}
\label{eq:dressed_op}
\left(\Theta_n\right)_{ij} (t) = (s_n)_{ij} \int_0^t d \tau e^{-i \omega_{ij} \tau} C_n (\tau)
\end{equation}
where $\omega_{ij} = (\epsilon_i - \epsilon_j)$ are the dimensionless frequencies of the system given by scaled differences in the eigenvalues, $\epsilon_i$, of $\tilde{H}_S$. The bath correlation function, $C_n (t)$ is given by
\begin{align}
\label{eq:bcf}
C_n (t) = \frac{1}{\pi} \int_0^{\infty} d \omega \, J_{n,\text{fast}} (\omega)& \left[ \coth (\beta  \omega /2) \cos (\omega t) \right. \nonumber \\
&- i \sin (\omega t) \left. \right].
\end{align}
where $\beta$ is inverse temperature times Boltzmann's constant. Since TCL2 stems from second-order perturbation theory, we expect for it to be accurate when $\eta \ll 1$, where the dimensionless coupling is computed over only the fast modes, $J_{n,\text{fast}} $, with a characteristic frequency given by $\omega^*$. Together, this hybrid formulation, denoted TCL2-FM, due to Montoya-Castillo, Berkelbach and Reichman,\cite{montoya2015extending} offers a potentially computationally efficient and accurate\cite{fetherolf2017linear,tempelaar2018vibronic} way to study motion through conical intersections under our physically motivated assumptions of scale separation. 
%
\section{\label{sec:results}Comparison with exact results}

\subsection{\label{sec:pyr}Non-markovian bath limit}
To understand the effectiveness of this approach, we first consider the case where the characteristic electronic timescale, $\lambda$, is well separated by the characteristic bath frequency, $\omega_c$, such that $\omega_c/\lambda \ll 1$. This is expected to hold when the remaining bath degrees of freedom are described by long wavelength solvent modes, either from slow dipolar or density fluctuations.\cite{song1996gaussian,fleming1986activated} We explore this regime in the relaxation of the $S_2(\pi\pi^*)-S_1(n\pi^*)$ conical intersection of pyrazine, following a model 
developed by Kuhl and Domcke.\cite{kuhl2000effect,kuhl2002multilevel} The Hamiltonian  has the form of a linear vibronic model, with an additional ground electronic state. In dimensionless harmonic-oscillator coordinates it is given by,
\begin{align}
H_S &=\ket{0} h_0 \bra{0} + \sum_{i,j=1,2} \ket{i} h_i \delta_{ij} + \lambda q_c(1-\delta_{ij})  \bra{j} \\ \nonumber 
h_i &= h_g + V^{(0)}_i + \kappa_t^{(i)} q_t \\ \nonumber
h_0 &=   \sum_{n=c,t} \frac{\Omega_k}{2} \left(-\frac{\partial^2 }{\partial q_k^2} + q_k^2 \right)
\end{align}
where $\Omega_{t(c)}$ is the frequency of the tuning (coupling) mode, $\kappa_t^{(i)}$ denotes the Holstein-like coupling of the tuning mode to each electronic state $i$, $h_0$ denotes the Hamiltonian of the ground electronic state, and the vertical energy shifts from the ground state are $V^{(0)}_i$'s. There are no other Holstein-like couplings, so the system-bath coupling is given by
\begin{equation}
H_{SB} = \left (\ket{1} \bra{1} + \ket{2} \bra{2} \right ) \sum_{n=c,t} q_n  \sum_{k} c_{n,k} Q_{n,k}.
\end{equation}
with spectral densities of the Debye form,
\begin{equation}
J_n (\omega) = 2 E_{r,n} \omega_{c,n} \frac{\omega}{\omega^2 + \omega_{c,n}^2}, \quad n=c,t,
\end{equation}
\begin{figure}[t]
\centering
\includegraphics[width=8.5cm]{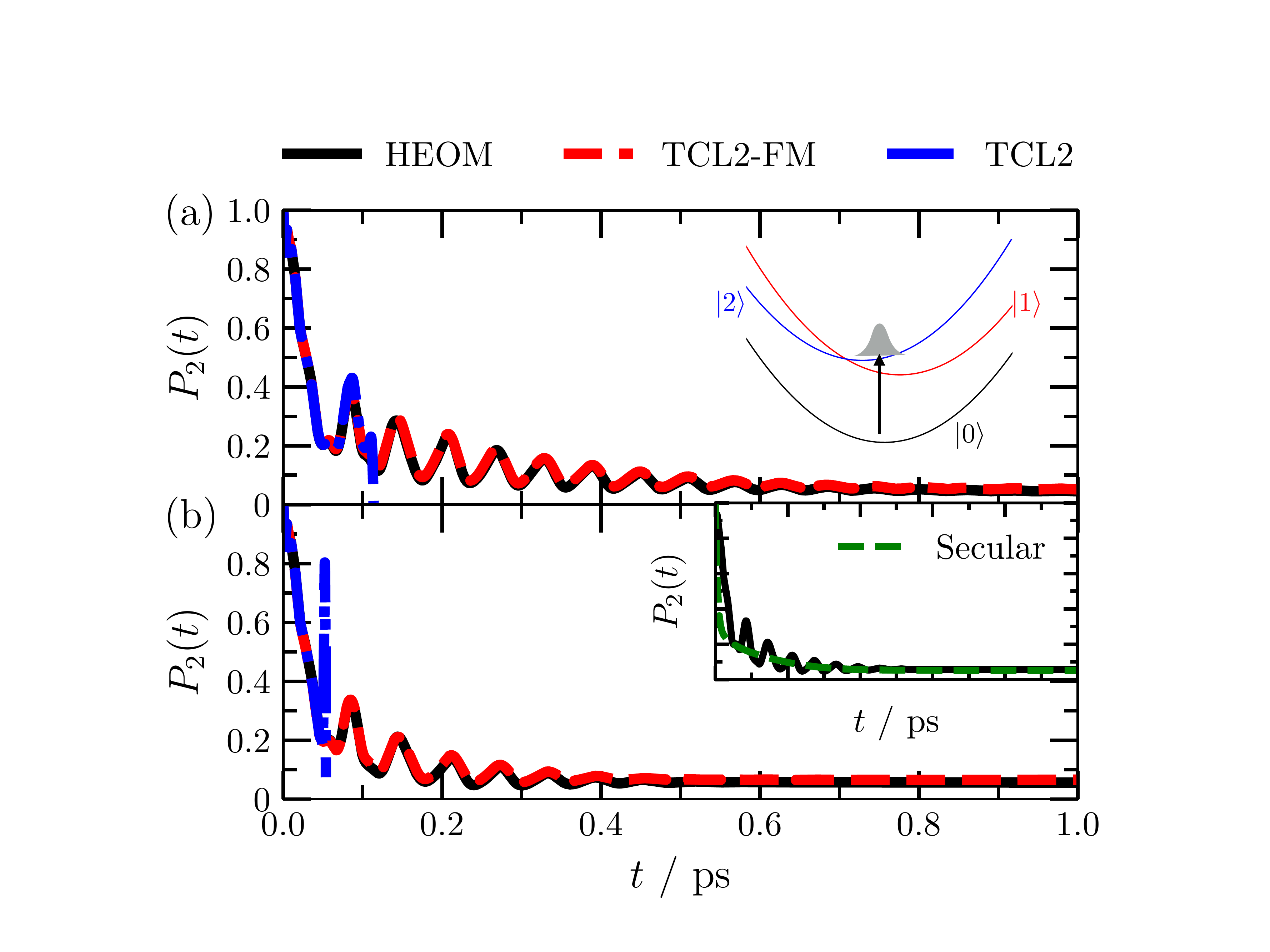}
\caption{Diabatic populations given by Eq. \ref{eq:pyr_p2} for a bath with $\omega_c = 0.0132$ eV with values of the reorganization energy $E_r = 0.006571$ eV (a) and $E_r = 0.01314$ eV (b). Shown in the inset of (a) are the potential energies for each electronic state along the coordinate $q_t$. In the inset of (b) is shown the results for secular Redfield theory with frozen modes. HEOM data was taken from Ref. \onlinecite{chen2016dissipative}.}
\label{fig:popstau50}
\end{figure}
which results from an exponentially decaying bath correlation function. The form of system bath coupling induces vibrational relaxation in each of the electronic states. The specific parameters for the system are $\Omega_c = 0.118 $, $\Omega_t = 0.074$, $\kappa_t^{(1)} = -0.105$, $\kappa_t^{(2)} = 0.149$, $\lambda = 0.262$, $V^{(0)}_1 = 3.94$, and $V^{(0)}_2 = 4.84$, all in eV, while the temperature of the bath was taken to be 300 K. The initial condition is generated by vertical excitation from the ground electronic state $\ket{0}$ into the diabatic electronic state $\ket{2}$ by
\begin{equation}
\sigma(0) = \ket{2} \ket{\chi_{02}} \bra{\chi_{02}} \bra{2},
\end{equation}
where $\ket{\chi_{02}}$ denotes the vibrationally-coherent wavepacket obtained from Frank-Condon overlaps between the ground electronic state $\ket{0}$ and electronic state $\ket{2}$. The system was expanded in a direct product basis of 20 harmonic oscillator eigenstates for each mode, making the system size 800 total states. The dynamics were propagated in a truncated basis, which with this initial condition is converged by considering only the lowest 500 energy eigenstates. A sketch of the system is shown in Fig.~\ref{fig:popstau50}(a).
\begin{figure}[b]
\centering
\includegraphics[width=8.5cm]{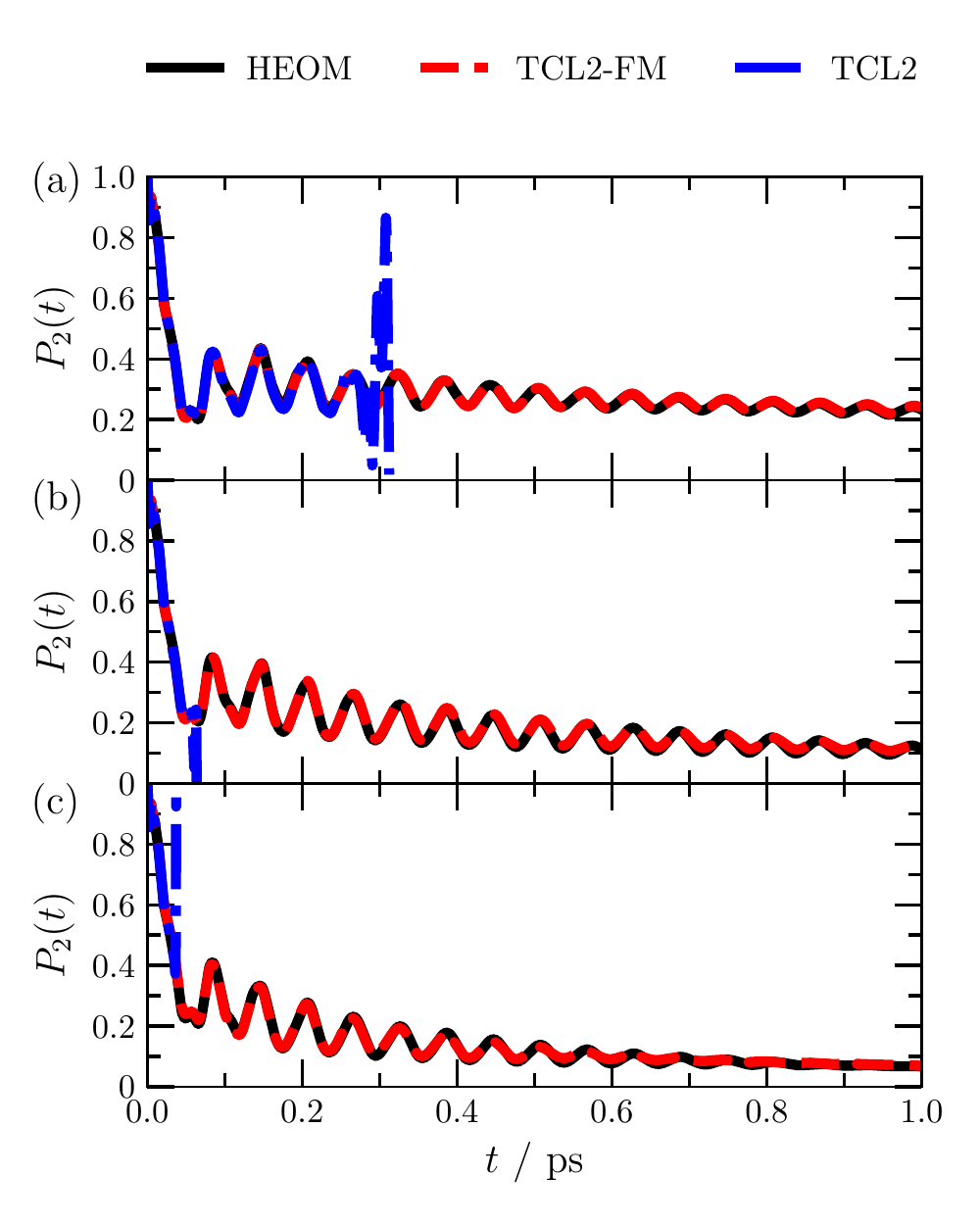}
\caption{Diabatic populations for a bath with  $\omega_c = 0.00397$ eV with values of the reorganization energy $E_r = 0.001314$ eV (a), $E_r =  0.006571$ eV (b), and $E_r = 0.01314$ eV (c). HEOM data was taken from Ref. \onlinecite{chen2016dissipative}.}
\label{fig:popstau166}
\end{figure}

We compare the validity of the dynamics obtained from TCL2 and the hybrid TCL2-FM, to the dynamics obtained from the numerically exact hierarchy equations of motion (HEOM) method  by Chen \textit{et al.}\cite{chen2016dissipative} These calculations were converged using the same basis with a hierarchy depth of 12. Since the system was at high temperature, no Matsubara terms were included. We first compute the time-dependent diabatic population in electronic state $\ket{2}$,
\begin{equation}
\label{eq:pyr_p2}
P_2 (t) = \tr \{ \ket{2}\bra{2} \sigma (t) \}.
\end{equation}
from a trace over all vibronic states. Two different characteristic frequencies of the bath are compared, a fast bath in which both the tuning and coupling modes are $\omega_{c,(c,t)} = 0.0132$ eV and a slower bath in which $\omega_{c,(c,t)} = 0.00397$ eV. 
Thus, in both cases the bath relaxes on a timescale of at least an order of magnitude slower compared to the Peierls coupling, $\omega /\lambda \ll 1$, and we can choose a large value of $\omega^*$ to treat the slow degrees of freedom. Details on the sensitivity of the results to the specific choice of $\omega^*$  are reported in the Appendix,  but over the range from $\omega^*=[0.0165,0.0329]$ we obtain nearly indistinguishable population dynamics. For both baths studied, we choose $\omega^* = 0.0219$ eV. Only 50 initial conditions are needed to obtain well-converged populations, which are drawn from a Boltzmann distribution with 1000 modes for each bath using the discretization procedure outlined in Ref. \onlinecite{montoya2015extending}.
\begin{figure*}[t]
\centering
\includegraphics[width=0.95\textwidth]{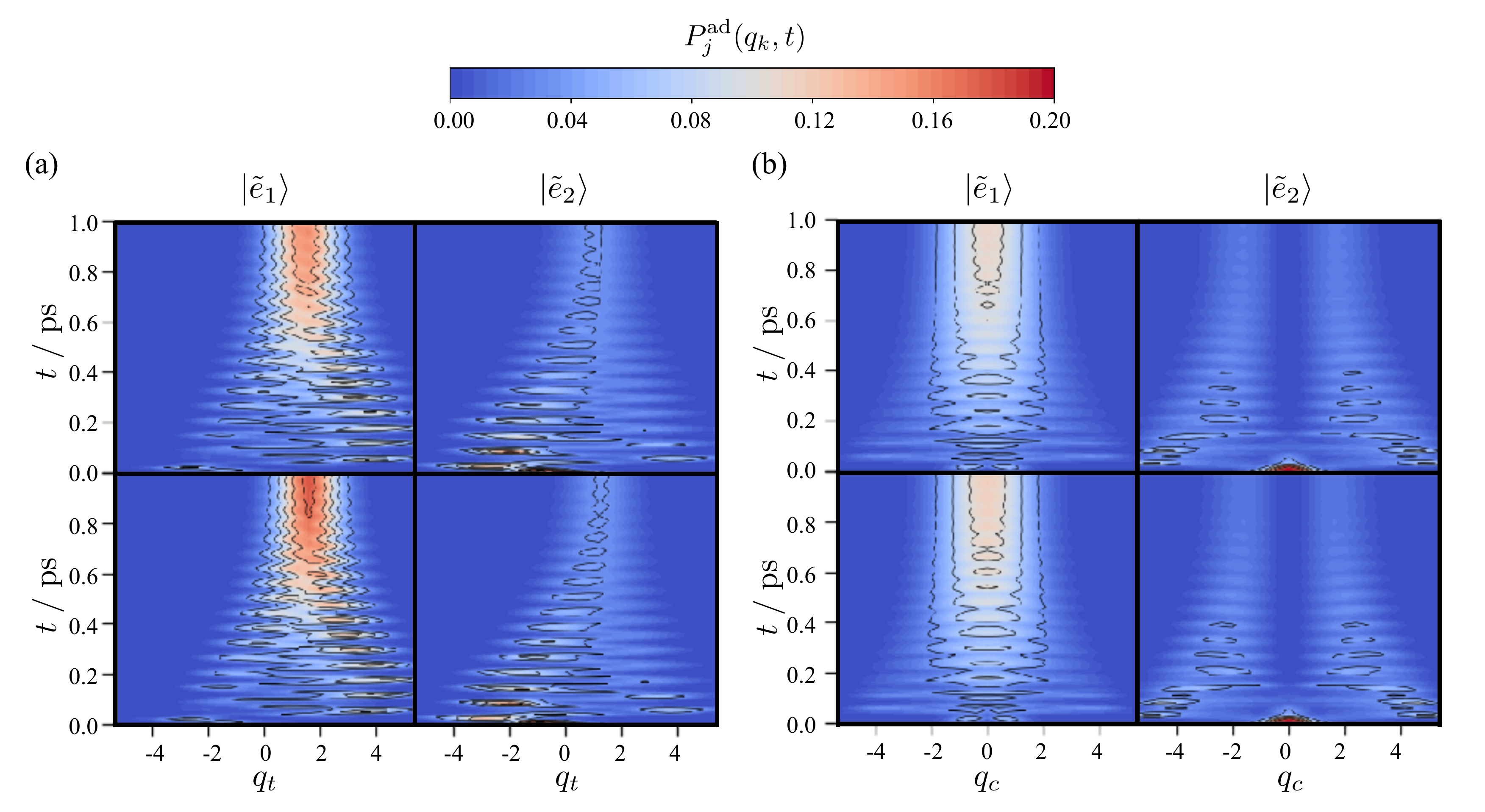}
\caption{Projections onto the adiabatic ground- (left column) and excited-state (right column) surfaces for the dimensionless coordinate $q_t$ (a) and for $q_c$ (b) for the pyrazine system with $\omega_c = 0.0132$ eV and $E_r = 0.006571$. The top row shows results from TCL2 -FM while the bottom row shows results from HEOM from Ref. \onlinecite{chen2016dissipative}.}
\label{fig:pyr_surfaces}
\end{figure*}

The populations obtained in the case of the faster bath are compared in Fig. \ref{fig:popstau50}. Ultrafast relaxation from state $\ket{2}$ into state $\ket{1}$ occurs within 50 fs, as the initial wavepacket proceeds through the conical intersection. This is followed by a prolonged period of coherent wavepacket motion that persists up to 0.5 ps, before decohering.
At weaker system-bath couplings than the ones presented here, TCL2 exhibits quantitative accuracy compared to HEOM. At larger system bath coupling strengths, $E_r = $ 0.006571 and 0.01314 eV, TCL2 exhibits positivity violations of the density matrix, which for fixed time step leads to instabilities in the dynamics. This failure is due to the breakdown of perturbation theory and requires contributions from higher-order correlation functions, as multiphonon processes become important. This is evident by noting that the dimensionless couplings are $\eta = 0.317$ and 0.634, which are not much less the 1 as required by perturbation theory.  

The hybrid approach, TCL2-FM, removes all positivity violations from TCL2 and achieves quantitative accuracy at all values of the reorganization energy studied, as compared to HEOM. The stability of the dynamics is a consequence of the frozen modes reducing the dimensionless couplings by nearly an order of magnitude, to $\eta = 0.055$ and 0.086, returning the treatment of the bath into the perturbative regime. The accuracy is a consequence of the small effect of the slow modes on the dynamics, acting only to further decohere vibrational oscillations but not significantly dissipate energy, due to the large separation of timescales between system and bath relaxation.

By invoking both the Markovian approximation, which takes the time integral in Eq. \ref{eq:dressed_op} to infinity, and the secular approximation, which decouples the dynamics of the populations from coherences in the energy eigenbasis\cite{breuer2002theory}, we get an equation of motion that is guaranteed to preserve positivity of the density matrix.\cite{lindblad1976generators,gorini1976completely} These approximations fail to exhibit the extended vibrational coherence and over-estimates the rate of relaxation. Neither effects are improved by the addition of frozen modes. The lack of vibrational dephasing is due to the neglect of coherence-coherence couplings in the relaxation tensor within the secular approximation\cite{kuhl2002multilevel} and the over-estimation of the rate is due to the Markovian approximation. As shown in the inset to Fig. \ref{fig:popstau50}b), they do, however, obtain the correct long-time of the populations as thermalization with the environment is accurately modeled.

Shown in Fig. \ref{fig:popstau166} are the populations for the case of the slower bath, where non-Markovian effects are more pronounced. As expected, TCL2 fails at an even smaller reorganization energy than in the fast bath regime due to the violation of the 2nd-order cumulant approximation. TCL2-FM remedies this failing and recovers quantitative accuracy for all reorganization energies studied. Again the stability is a consequence of reducing the dimensionless coupling by an order of magnitude. In this case the original couplings are 
$\eta=$ 0.210, 1.05 and 2.10, and are reduced to $\eta=$ 0.021, 0.106 and 0.213 by freezing the slow modes.  As has been noted previously,\cite{montoya2015extending} the inclusion of the slow modes as static disorder effectively incorporates all-order effects from those bath modes, albeit only their influence on the altered eigen-structure of the Hamiltonian. When the timescales of system and bath relaxation are well separated, as is expected to hold generally in systems with conical intersections where electronic relaxation is ultrafast, this frozen mode approximation allows for an accurate low order quantum master equation description of the dynamics. 

A rigorous test of the accuracy of frozen modes can be obtained by comparing the projection of the wavepackets in the adiabatic basis obtained from, 
\begin{align}
P_j^{\text{ad}} (q_t, t) &= \int d q_c \bra{q_c} \bra{q_t} \bra{\tilde{e}_j} \sigma (t) \ket{\tilde{e}_j} \ket{q_t} \ket{q_c} \\
P_j^{\text{ad}} (q_c, t) &= \int d q_t \bra{q_c} \bra{q_t} \bra{\tilde{e}_j} \sigma (t) \ket{\tilde{e}_j} \ket{q_t} \ket{q_c} 
\end{align}
where $\ket{\tilde{e}_{j=1,2}}$ are the adiabatic electronic wavefunctions given by the diabatic-to-adiabatic transformation\cite{manthe1990dynamics} 
\begin{equation}
\ket{\tilde{e}_{j}} = \sum_{j'} S(q_c, q_t) \ket{j'}
\end{equation}
where $S(q_c, q_t)$ is the rotation matrix given by
\begin{equation}
S(q_c, q_t) = \begin{pmatrix} \cos \alpha (q_c, q_t) & -\sin \alpha (q_c, q_t) \\ \sin \alpha (q_c, q_t) & \cos \alpha (q_c, q_t) \end{pmatrix}
\end{equation}
and $\alpha (q_c, q_t)$ is the diabatic-to-adiabatic mixing angle. These projections record information about the entire density matrix since it requires unitary transformations acting on both populations and coherences. Figure~\ref{fig:pyr_surfaces} (a) shows the projection of the wavepacket along the tuning mode obtained from TCL2-FM compared to those obtained from HEOM. The results from the TCL2-FM approach are virtually indistinguishable from the HEOM results at all times. This implies that the full density matrix is accurately computed with TCL2-FM. Projections along the coupling mode are shown in Fig.~\ref{fig:pyr_surfaces} (b). Again, TCL2-FM exhibits quantitative accuracy. That the full density matrix is accurately obtained also implies that arbitrary observables, including spectroscopic signals\cite{fetherolf2017linear} might be reliably computed.

\subsection{Markovian bath limit}
\begin{figure}[t]
\centering
\includegraphics[width=8.5cm]{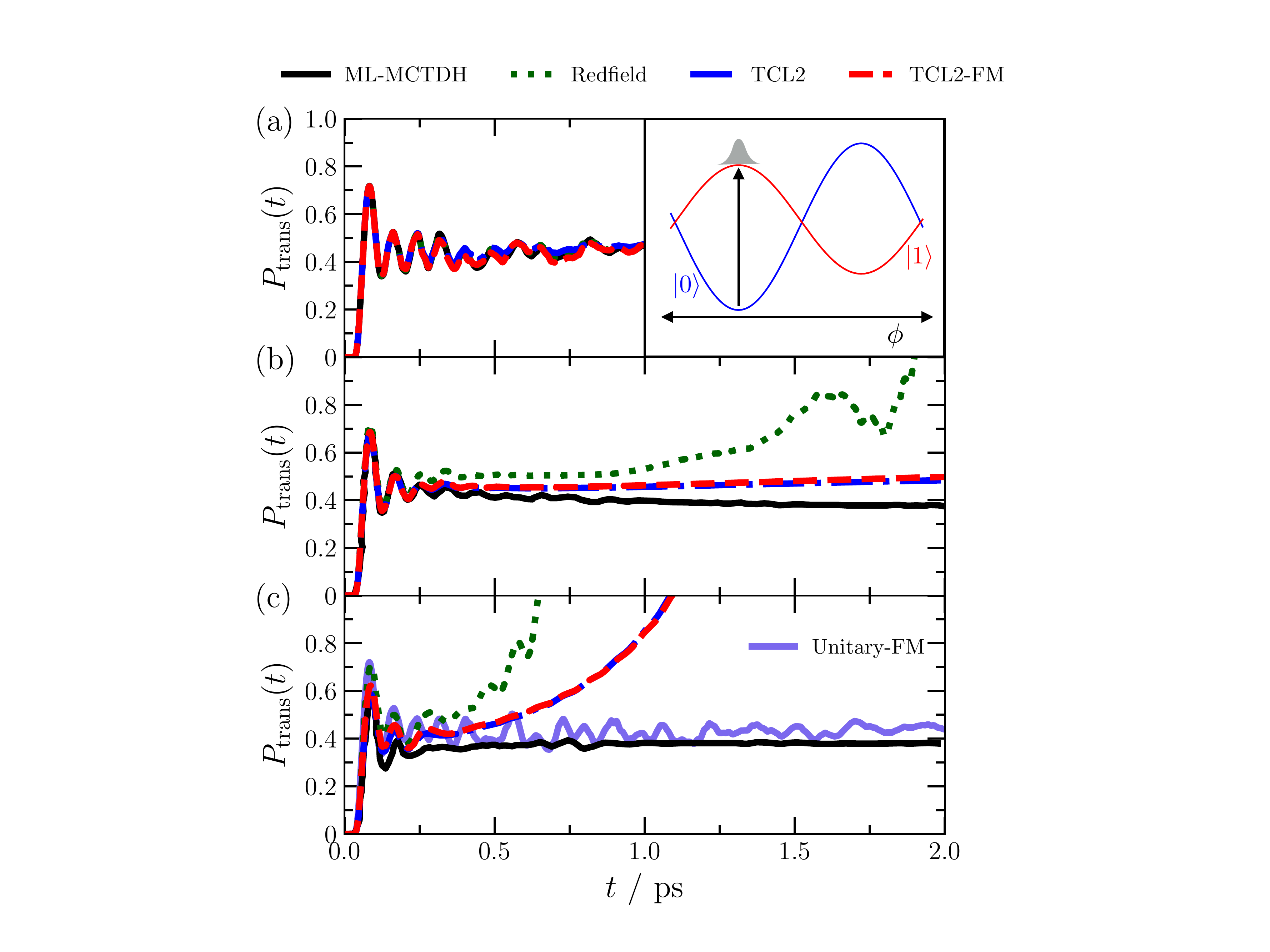}
\caption{Shown are the results for the population in the trans state at different reorganization energies: $E_r = 0.0159$ eV (a), $E_r = 0.159$ eV (b), and $E_r = 0.318$ eV (c). Numerically exact ML-MCTDH results from Ref. \onlinecite{thoss2006quantum} are shown in dashed black lines, Redfield theory in green dotted, TCL2 in blue dashed-dotted lines, and TCL2 -FM in dashed red. The inset of (a) shows a figure of the periodic system along the $\phi$ coordinate. In (c) we also show the results when the entire bath is discretized and frozen (solid purple) giving rise to a purely unitary dynamics for each realization of bath modes. The unitary-FM dynamics were obtained by sampling over 100 trajectories.}
\label{fig:thoss_ptrans}
\end{figure}

To understand the limits of this approach, we next consider the case where the characteristic electronic timescale, $\lambda$, is not separated by the characteristic bath frequency, $\omega_c$, such that $\omega_c/\lambda \sim 1$. This limit is expected when the remaining bath degrees of freedom couple directly to the electronic states through optical solvent modes or to high frequency vibrations. We study this case in a model for the photoisomerization dynamics of retinal rhodopsin, shown in the inset of Fig. \ref{fig:thoss_ptrans} (a). This model has been studied by Thoss and Wang using the numerically exact multilayer formulation of MCDTH, ML-MCTDH.\cite{thoss2006quantum} The model describes the dynamics along a periodic isomerization coordinate, $\phi$, which plays the role of the tuning mode, and a harmonic coupling coordinate, $q_c$. 

The system Hamiltonian has the following form,
\begin{equation}
H_S = \sum_{i,j=0,1} \ket{i} (T + V_{i})\delta_{ij} + \lambda q_c (1-\delta_{ij})  \bra{j}, 
\end{equation}
where $T$ is the total kinetic energy operator,
\begin{equation}
T = -\frac{1}{2I} \frac{\partial^2}{\partial \phi^2} - \frac{\Omega_c}{2} \frac{\partial^2}{\partial q_c^2},
\end{equation}
where  $I$ is the moment of inertia for the tuning mode. The potential energies for each electronic state, $V_{i}$, are
\begin{align}
V_{i} &= V^{(0)}_i + (-1)^i \frac{1}{2} W_i (1-\cos \phi) + \frac{\Omega_c}{2} q_c^2 + \delta_{1i} \kappa_c q_c 
\end{align}
where $W_n$ are the energy amplitudes of the isomerization potential, and $V_i^{(0)}$ are the energy shifts of each diabatic state relative to the energy in the \emph{cis} state. The coupling mode is described by the frequency $\Omega_c$ and Holstein coupling $\kappa_c$. The specific parameters for this model are $I^{-1} = 1.43 \times 10^{-3}$, $V^{(0)}_0 = 0.0$, $V^{(0)}_1 = 2.0$, $W_0 = 2.3$, $W_1 = 1.5$, $\Omega_c = 0.19$, $\lambda = 0.19$, and $\kappa_c = 0.095$, all in eV. The system was expanded in a basis of plane waves for the isomerization mode and harmonic oscillator eigen- states for the coupling mode with a basis set size of 301 and 24, respectively. This choice gave a Hilbert space size of 14448 states, but the dynamics were converged using only the lowest 1000 energy eigenstates.

The form of the system-bath coupling is given by
\begin{equation}
H_{SB} = \ket{1}\bra{1} \sum_k c_k Q_k,
\end{equation}
which describes the response of a polar solvent to an instantaneous change in the charge distribution of the molecule. The spectral density used is Ohmic with an exponential cutoff,
\begin{equation}
J (\omega) = \frac{\pi E_r}{\omega_c} \omega e^{-\omega / \omega_c},
\end{equation}
and the value of this cutoff frequency used was $\omega_c = 0.2$ eV.
The temperature was taken to be 0 K. The initial condition was a vertical excitation of the  ground vibrational state of electronic state $\ket{0}$ into electronic state $\ket{1}$, given by
\begin{equation}
\sigma (0) = \ket{1} \ket{\chi_{01}} \bra{\chi_{01}} \bra{1},
\end{equation}
where again $\ket{\chi_{01}}$ denotes the vibrationally-coherent wavepacket obtained from Frank-Condon overlaps between the two electronic states.

To test the validity of the TCL2 with frozen mode approach, we simulated the dynamics up to 2 ps and compared to the exact result obtained from ML-MCTDH for a range of reorganization energies, which represented the system degrees of freedom and a discretized bath of $\sim40$ modes explicitly. We specifically compute the time-dependent population of the  \emph{trans} state,
\begin{equation}
P_{\text{trans}} (t) = \tr \{ \theta ( | \phi | - \pi /2 ) \},
\end{equation}
where $\theta(x)$ is the Heaviside step function and the trace implies integration over the $\phi$ and $q_c$ coordinates, following initial excitation. For this model, the electronic timescale inferred from the Peierls coupling, $\lambda$ is nearly the same as the characteristic frequency of the bath, $\omega_c$, or $\omega_c/\lambda \sim 1$. Since the electronic and bath timescales are not well separated, we expect that while choosing to freeze some modes of the bath will reduce the system-bath coupling and stabilize the perturbation theory description of the fast bath modes, this will come at a cost of incorrectly describing the time-dependent dissipation as modes that are being held frozen should contribute. 
We first consider the consequences of choosing $\omega^* = \omega_c$, which will reduce the strength of coupling from modes that have frequencies smaller than the position of the peak in the spectral density, while treating the peak and modes with higher frequency with perturbation theory.
We discretized the bath using 1000 modes and sampled over the Wigner transform of the Boltzmann distribution. Only five trajectories were averaged over due to the negligible effect of the frozen modes to TCL2 dynamics as discussed below.

Figure~\ref{fig:thoss_ptrans} shows the time dependent population in the \emph{trans} state. At the smallest value of the reorganization energy used, $E_r = 0.0159$ eV, shown in Fig.~\ref{fig:thoss_ptrans}(a), the dynamics are characterized by relaxation of the population after 0.1 ps and highly damped decay of vibrational coherences on a similar timescale.  For this case, Markovian Redfield theory and TCL2 are nearly indistinguishable. This is a consequence of being well within the weak coupling limit, with $\eta=0.050$. The dynamics are in quantitative agreement with available numerically exact ML-MCTDH results.\cite{thoss_results} Adding frozen modes has no real effect on the dynamics, which might be expected at a small value of system-bath coupling. 

At a reorganization energy that is a factor of ten larger, $E_r = 0.159$ eV, Redfield theory exhibits positivity violations that render the dynamics unstable after 1.5 ps. These results are shown in  Fig.~\ref{fig:thoss_ptrans}(b). These violations are corrected by TCL2, over short times, but at longer times TCL2 also becomes unstable. Using frozen modes stabilizes the dynamics at longer times, but has no effect at intermediate times. In all three descriptions the early qualitative features are correct, but the population in the \emph{trans} state is too large for TCL2 and TCL2-FM at 2 ps and it does not decrease at long times as the exact ML-MCTDH results do. Under these conditions, the coupling to the bath is reduced from $\eta = 0.495$ to $\eta=0.177$ using frozen modes, which are still larger than should be expected to yield accurate results. Thus, there are expected multi-phonon processes that are missed by the perturbative treatment in TCL2. 

At even larger reorganization energies, both TCL2 and TCL2-FM show positivity violations and result in unstable dynamics past 1 ps. These results are shown in  Fig.~\ref{fig:thoss_ptrans}(c). At this value of the reorganization energy, the couplings to the bath both without and with frozen modes are $\eta=1.01$ and $\eta=0.372$ respectively, are too large to self-consistently truncate the cumulant expansion at second order. By taking $\omega^*$ to be larger, we can sufficiently reduce the coupling to the remaining bath degrees of freedom that the dynamics are stable, but still dissipative within a TCL2 description. 
However, the dynamics deviate from the numerically exact result, as the approximation that degrees of freedom with $\omega<\omega^*$ are static, is not valid as $\lambda<\omega^*$, leading to a description of the dynamics that is not consistent. While in the pyrazine model the large separation of time scales allowed a large range of $\omega^*$ to be selected without disrupting the subsequent relaxation dynamics, this separation is not present for the rhodopsin model studied. 
\begin{figure*}[t]
\centering
\includegraphics[width=0.85\textwidth]{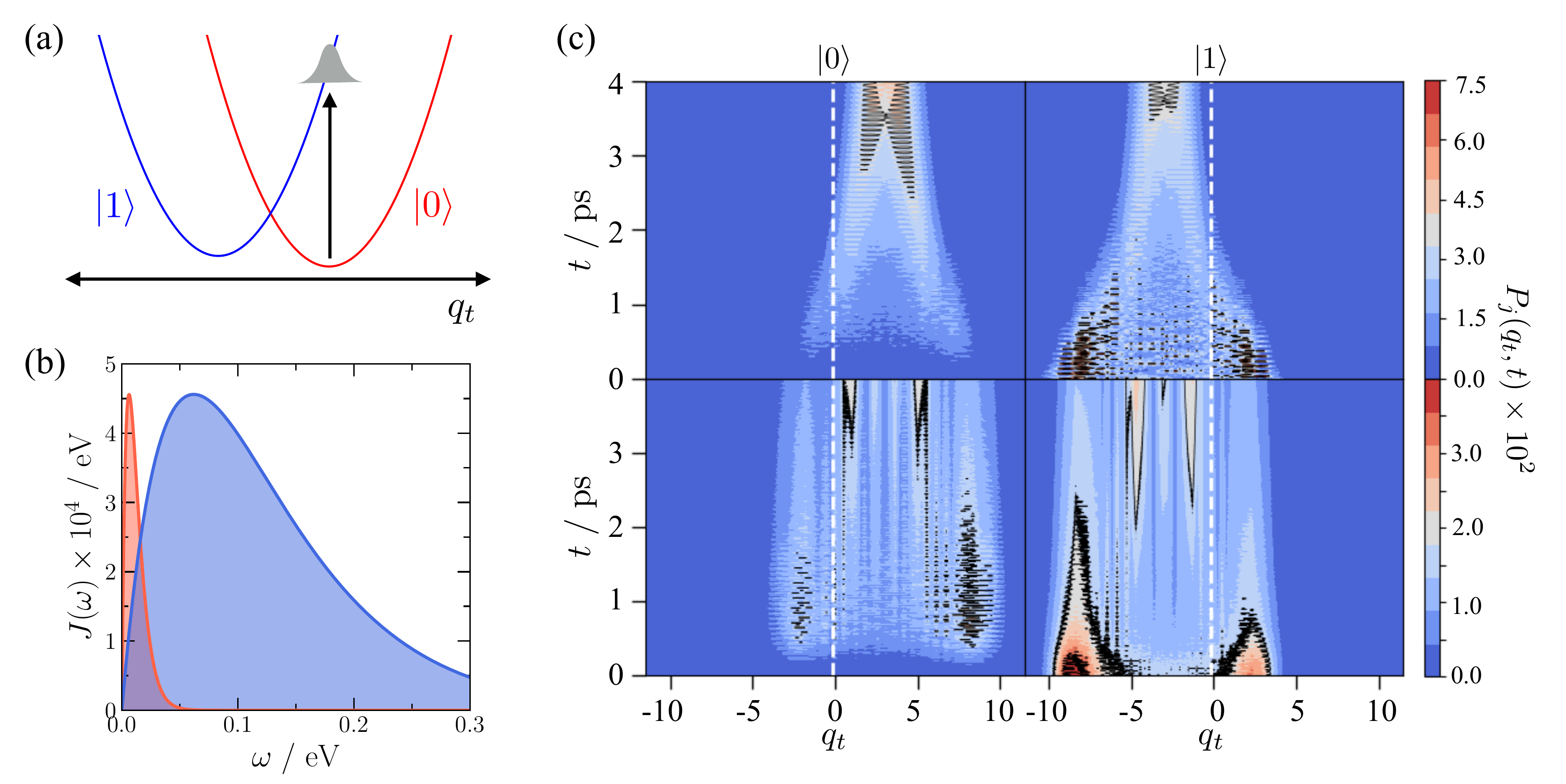}
\caption{Shown in (a) is the harmonic oscillator model with initial condition. The spectral densities used are shown in (b) with $\omega_c = \Omega_t$ in blue and $\omega_c = \Omega_t /10$ in red. Projections onto the diabatic electronic surfaces in the dimensionless coordinate $q_t$ are shown in (c) for $E_{r,t} = 0.2$ with $\omega_c = \Omega_t$ (top) and $\omega_c = \Omega_t /10$ (bottom).}
\label{fig:meta_surf}
\end{figure*}

To formulate a correct description of the system dynamics in the limit of strong system-bath coupling when motion in the system and bath are on similar time scales requires that the reorganization energy be reduced without freezing fast bath modes. This could be done by adding an additional effective bath mode into the description of the system, whose dynamics would be treated exactly. While including additional modes into the system Hamiltonian dramatically increases the Hilbert space, using unravelling techniques that reduce the scaling of master equation propagation\cite{kleinekathofer2002stochastic,vogt2013stochastic}, adding a few additional modes is possible. This is an active area of research, though beyond the scope of the present study.

\section{Application to Photoisomerization Quantum Yields}
\label{Sec:Yields}
With the limitations of our approach mapped out, we now study the dependence of the photoisomerization quantum yield on the bath. We consider the nonadiabatic relaxation through a conical intersection of a linear vibronic model constructed to have features similar to those in a molecular photoisomerization processes.\cite{ikeda2019modeling,gruber2019action} Specifically, we construct a model where a conical intersection lies above two adjacent basins, one metastable with respect to the other. Our approach enables us to study a wide range of system-bath coupling strengths in the Markovian and non-Markovian regimes and understand the impact of the yield on these parameters. Describing the dynamical features arising in such complex environments are paramount to describing the yields, as they are completely determined by relaxation rates rather than being constrained by thermodynamic considerations.\cite{yunger2018fundamental}

The Hamiltonian we consider has the form,
\begin{align}
H_S &= \sum_{i,j=0,1} \ket{i} h_i \delta_{ij}  + \lambda q_c (1-\delta_{ij}) \bra{j}  \\ \nonumber
h_i &= \sum_{k=c,t} \frac{\Omega_k}{2} \left(-\frac{\partial^2}{\partial q_k^2}+ q_k^2 \right) + \kappa_t^{(i)} q_t + V^{(0)}_i
\end{align}
with a system-bath coupling,
\begin{equation}
H_{SB} = (\ket{0} \bra{0} + \ket{1} \bra{1} ) \sum_{n=c,t} q_n  \sum_{k} c_{k,n} Q_{k},
\end{equation}
meant to model vibrational relaxation and an Ohmic spectral density with exponential cutoff,
\begin{equation}
J_n (\omega) = \frac{\pi E_{r,n}}{\omega_{c,n}} \omega e^{-\omega / \omega_{c,n}}, \quad n=c,t.
\end{equation}
for both the coupling and tuning modes. The model described here, shown in Fig. \ref{fig:meta_surf} (a), is similar to a model studied by Thorwart and co-workers\cite{qi2017tracking,duan2018signature}. We set the parameters to be $\Omega_c = 0.112 $, $\Omega_t = 0.0620$, $\kappa_t^{(0)} = -0.186$, $\kappa_t^{(1)} = 0.186$, $\lambda = 0.0248$, $V^{(0)}_1 = -0.031$, and $V^{(0)}_2 = 0.031$ in eV, while the temperature of the bath is taken to be 300 K. The Hamiltonian was expanded in a basis of harmonic oscillator eigenstates with 75 states used for the tuning mode and 5 states for the coupling mode. The dynamics were propagated in the energy eigenbasis with a truncated basis of 400 states, which shows convergence to the full Hilbert space.

We have tuned the system Hamiltonian parameters to include a metastable well in the higher-energy electronic state. The barrier to transferring population along the ground adiabatic state is $\sim 0.129$ eV, so there will be a separation of timescales between initial relaxation into the minima of the two diabatic states and subsequent barrier crossings. 
We consider the dynamics following a vertical excitation into state $\ket{1}$ from the ground vibrational state of $\ket{0}$,
\begin{equation}
\sigma (0) = \ket{1} \ket{\chi_{01}} \bra{\chi_{01}} \bra{1}.
\end{equation}
and are interested in the quantum yield into state $\ket{1}$ following subsequent relaxation over times long relative to vibrational relaxation, but short relative to relaxation into a thermal state.

We have studied the dynamics of this model with two different environments, one in the Markovian regime where $\omega_{c} \sim \lambda$, and one in the non-Markovian regime where $\omega_{c} \ll \lambda$. These two regimes are illustrated by their spectral densities in Fig. \ref{fig:meta_surf} (b). For both baths, we have studied the dynamics over a range of reorganization energies. The fast bath we study has a cutoff frequency of $\omega_{c,t} = 0.062$ eV for the tuning mode and $\omega_{c,c} = 0.112$ eV for the coupling mode. Since the bath is moderately fast relative to the timescale induced by the electronic coupling and the reorganization energies used are small, the largest has a coupling constant of $\eta < 0.1$, these populations are accurately obtained from TCL2 without the use of frozen modes. The slow bath we study has cutoff frequencies for the tuning mode $\omega_{c,t} = 0.0062$ eV with the coupling mode held fixed. Since this system is in a more non-Markovian regime, the dynamics using TCL2 alone exhibit positivity violations at significantly smaller values of the reorganization energy relative to the fast bath, and we thus use the frozen mode approach. However, we found it necessary to only freeze modes in the bath associated with the tuning mode. For each value of the reorganization energy we used $\omega^* = 0.00868$ eV, decreasing the largest value of the coupling to $\eta=0.05$. We find that choosing $\omega^*$ between 0.008 eV and 0.014 eV results in quantitatively similar population dynamics for all system bath coupling strengths considered. We simulated the dynamics with 50 trajectories, by discretizing the slow bath into 1000 modes for the tuning mode.

Shown in Fig. \ref{fig:meta_surf}(c) are the projections of the wavepacket onto the position basis of the tuning mode for each diabatic state, given by,
\begin{align}
P_j (q_t, t) &= \int d q_c \bra{q_c} \bra{q_t} \bra{j} \sigma (t) \ket{j} \ket{q_t} \ket{q_c} \, ,
\end{align}
for both baths. In the fast bath case, the wavepacket starts in electronic state $\ket{1}$ and coherently oscillates with enough energy to put it back in the Franck-Condon region at short times. The bath dissipates energy from this wavepacket, which reduces the vibrational coherence until the wavepacket can no longer reach the Franck-Condon region. In the slow bath case, the wavepacket dynamics are markedly different, showing an extended lifetime in higher-energy vibrational states. The rate of decoherence appears to be much faster as the oscillations of the wavepacket are damped out almost instantly, which is a reflection of the role of slow bath being a source of inhomogeneous broadening.

\begin{figure}[t]
\centering
\includegraphics[width=8.cm]{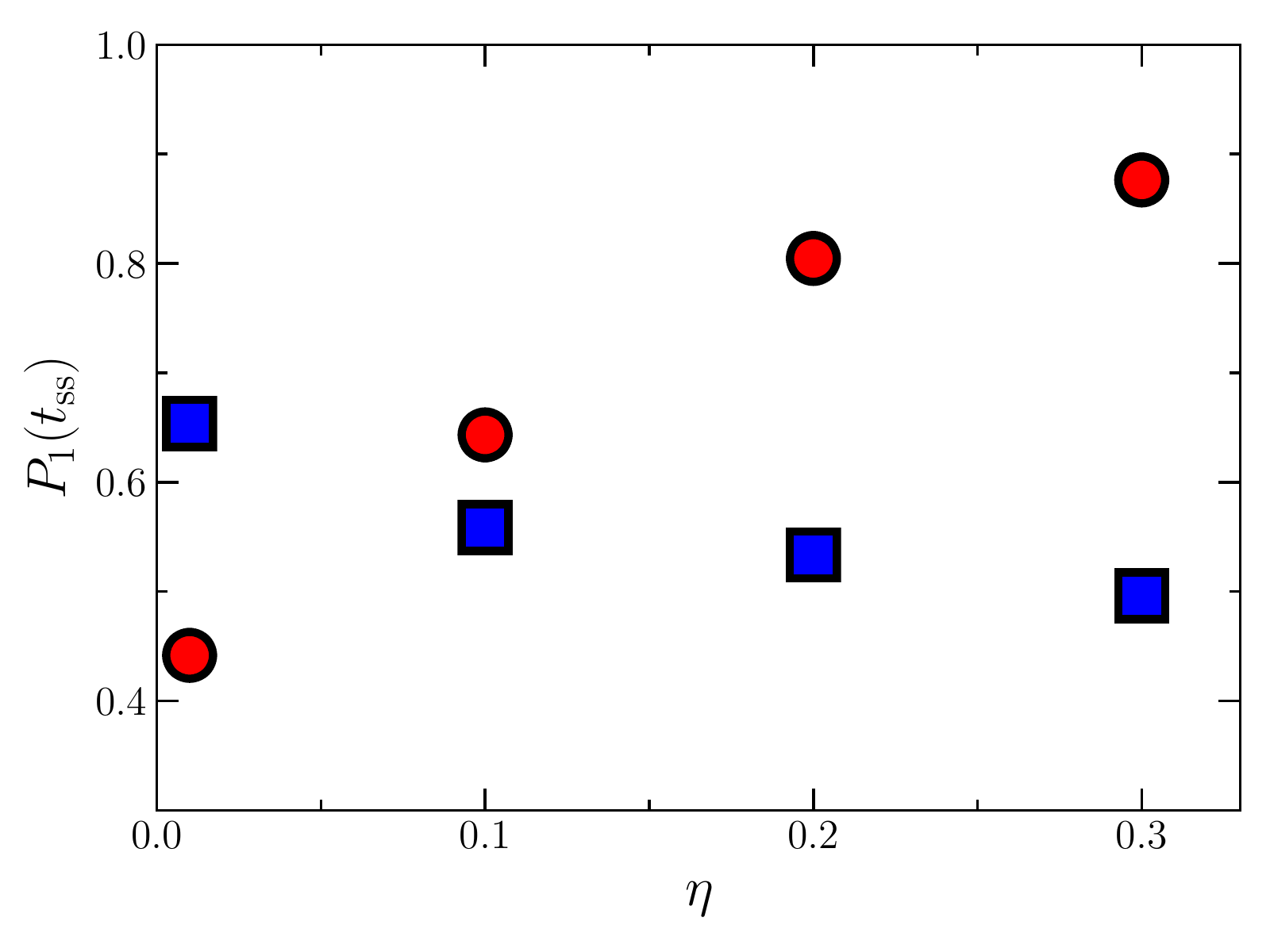}
\caption{Yields of $P_1 (t)$ taken in the quasi-steady-state limit for $\omega_{c,t} = \Omega_c/ 10$ (blue squares) and $\omega_{c,t} = \Omega_t$ (red circles), as a function of the coupling strength, $\eta$  computed from the total spectral density.}
\label{fig:meta_yields}
\end{figure}

These different relaxation mechanisms result in different quantum yields, and strikingly different dependence on the bath reorganization energy. We define the quantum yield as the diabatic population in state $\ket{1}$ in the quasi-steady-state limit,
\begin{equation}
P_1 (t_{\text{ss}}) = \tr \{ \ket{1}\bra{1} \sigma (t_{\text{ss}}) \},
\end{equation}
where $t_{\text{ss}}$ is the time taken for the diabatic populations to be nearly time invariant, which for the parameters studied is around 4 ps. In the case of the fast bath, we find the yield increases monotonically with the reorganization energy. This is shown in Fig. \ref{fig:meta_yields}, where $\eta$ is proportional to the reorganization energy with $\omega_c$ fixed and we take $\eta$ and $E_r$ from the total spectral density, not the reduced values from just the fast modes. The increase in the yield with reorganization energy in the fast bath is attributable to the fact that with increasing $E_r$, the wavepacket spends less time in the Franck-Condon region where population can transfer between the two diabatic states through electronic coupling. As is evident from the wavepacket dynamics, increasing the reorganization energy will increase the rate of vibrational dissipation and hence the localization of the wavepacket into the minima of the diabatic states. In the case of the slow bath, we find the opposite trend. Increasing the reorganization energy results in a decreasing the quantum yield. This decrease is attributable to the increased rate of decoherence and slower rate of dissipation due to the lag in the bath's ability to remove energy from the system. 

These results are in contrast to some other observations on related linear vibronic models. Previously, Thorwart and coworkers have found that the lifetime of vibrational coherence could be tuned by the reorganization energy or characteristic frequency of the bath and the persistence of this coherence had large impact on the photoisomerization yield.\cite{qi2017tracking,duan2018signature,duan2016impact} While we note that the former of these claims is verified by our simulations, we note that the diabatic potentials they studied do not have metastability as the zero point energy in the higher-energy electronic state is on the order of the barrier height for population transfer along the ground adiabatic state. Having a well-defined quantum yield, in which there is a long-lived metastable state, requires a separation of timescales between the initial relaxation and eventual thermalization. If the barrier height is not sufficiently large, as in their previous work,\cite{qi2017tracking,duan2018signature,duan2016impact} then there will not be a separation of timescales, and thus there will not be a uniquely defined quantum yield. Nevertheless, the complex dependence of the quantum yield on the parameters of the bath that we have found illustrates the rich chemical dynamics of conical intersection models that can be interrogated efficiently.

\section{\label{sec:conclusions}Conclusion}

In this paper, we have developed a strategy for simulating nonadiabatic relaxation through conical intersections in the condensed phase. The framework leverages the separation of timescales between the ultrafast dynamics of a few strongly coupled nuclear degrees of freedom, and the remaining weakly coupled degrees of freedom. In cases where the characteristic timescales of these two motions are well separated, when the nonadiabatic coupling is much larger than the characteristic frequency of the bath, we can consider the slowest of those modes frozen and treat the remaining with perturbation theory. Freezing the slowest modes produces a source of static disorder, and acts to decohere the resultant dynamics. Weak coupling perturbation theory, in the form of TCL2, correctly describes the time dependent dissipation to the environment and without the low frequency modes has a large domain of applicability. This is consistent with what has been previously observed in the context of the spin boson model.\cite{montoya2015extending}

When applicable, the strategy we have presented represents a computationally efficient framework for simulating dynamics in condensed phase environments. This efficiency is due to the optimal representation of subsets of degrees of freedom. The hybrid method used in this paper formally scales as $\mathcal{O} (t N_{\text{traj}} N_{\text{bath}} N^3)$ where $N_{\text{traj}}$ is the number of trajectories used that can be trivially parallelized, $N_{\text{bath}}$ is the number of baths and $N$ is the number of states in the system, and is linear in time, $t$.\cite{pollard1994solution} Wavefunction based methods like ML-MCTDH suffer from an exponential scaling in the size of the system that must be represented, albeit with a reduced scaling than naive direct product wavefunctions. This scaling arises in condensed phase models through the representation of explicit bath degrees of freedom, which causes super-linear scaling in the number of baths and causes exponential scaling in time due to the difficulty in avoiding Poincare recurrences. While exact quantum master equations, like HEOM, do not suffer from exponential scaling in time, they offer little benefit to the overall scaling as they scale factorially in in the number of auxiliary degrees of freedom that must be represented. This scaling causes significant memory requirements and also has super-linear scaling in the number of baths. This makes low temperature, and non-Markovian systems particularly difficult to study. As this approach extends the limitations of weak-coupling theories, it can be combined with importance sampling tools developed at weak-coupling to study reaction mechisms.\cite{schile2018studying} In molecular systems, when the number of degrees of freedom as well as anharmonicities in the system increases, we thus expect the hybrid approach of this paper to be useful in providing numerically accurate results.

\begin{appendices}
\section{Choosing $\omega^*$}

Here we consider the choice of the parameter $\omega^*$ in the TCL2-FM method for conical intersection models. We note that for site-exciton models an efficient choice has been found that partitions the bath based on the comparison of the Rabi frequencies and the characteristic frequency of the bath.\cite{montoya2015extending} Our discussion on this choice for conical intersection models will be \textit{ad hoc}, in that there will be no rigorously derived equation, but will provide a physically motivated procedure using the pyrazine model as an example. 
\begin{figure}[t]
\centering
\includegraphics[width=8.5cm]{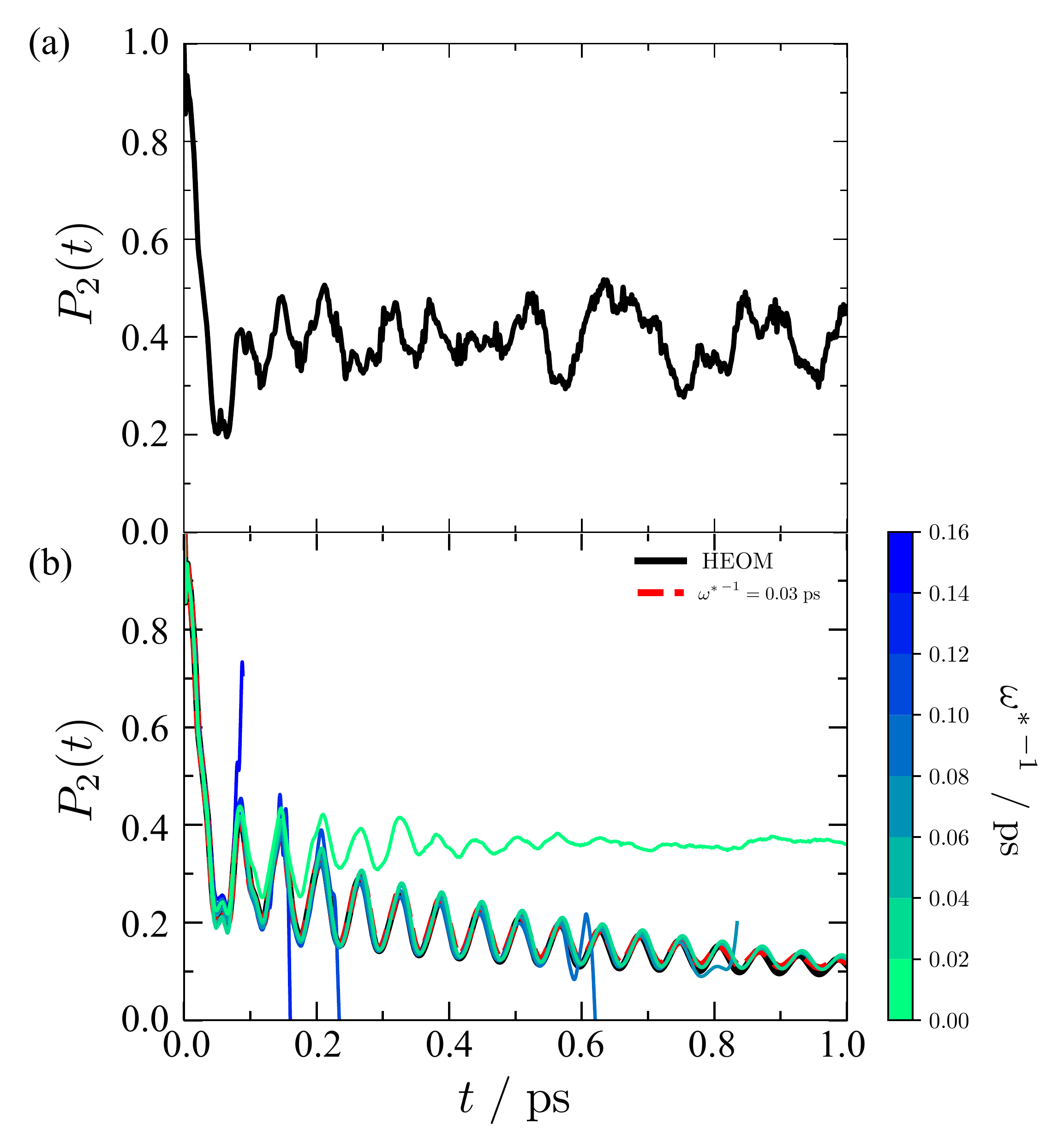}
\caption{Shown are the population dynamics for the pyrazine model without the presence of a bath (a). TCL2-FM with different values of $\omega^*$ corresponding to a timescale given in the colorbar are shown in (b). Also shown in (b) are exact HEOM  (solid black) and TCL2-FM (dashed red) results with $\omega^* = 0.0219$ eV from the main text. The bath parameters used were $\omega_c = 0.00397$ eV and $E_r = 0.006571$ eV.}
\label{fig:wstar_fig}
\end{figure}

Fig. \ref{fig:wstar_fig} (a) shows the dynamics of the pyrazine model without the presence of a bath. Within the first 30 fs, there is significant population transfer from electronic state $\ket{2}$ to state $\ket{1}$ followed by electronic beating that is modulated by the vibrational levels. These observations match those of Kr\v cm\'ar \textit{et al}, who compute two-dimensional electronic spectra in the two-mode pyrazine model with phenomenological dephasing.\cite{krvcmavr2014signatures} The spectra showed rapid population transfer between the two electronic states within 50 fs in addition to a complicated vibronic structure. The complex structure of this beating makes choosing a characteristic timescale of the system that can delineate between the slow and fast portions of the bath difficult. Despite this complexity, we infer that this first population transfer determines the splitting frequency for the bath. 

This hypothesis can be numerically tested by varying $\omega^*$ to treat less and less of the bath with TCL2 and incorporate more of the bath into the frozen modes description. Example diabatic populations are shown in Fig. \ref{fig:wstar_fig} (b) for a range of values of $\omega^*$ compared to the HEOM result and the TCL2-FM result from Fig. \ref{fig:popstau166} (b). For very small values of $\omega^*$ positivity violations are observed. As $\omega^*$ is increased these positivity violations become delayed until eventually they are washed out entirely. At values of $\omega^*$ corresponding to the range [0.0165,0.0329] eV, the numerically exact result is essentially reproduced, however, as $\omega^*$ is increased to infinity, so that the entire bath is treated as static, the results exhibit deviations due to the lack of dissipation in the completely static bath limit. 

\end{appendices}
\begin{acknowledgments}
The authors would like to acknowledge Prof. Michael Thoss, Prof. Eran Rabani, and Dr. Lipeng Chen for helpful discussions. This material is based upon work supported by the U.S. Department of Energy, Office of Science, Office of Advanced Scientific Computing Research, Scientific Discovery through Advanced Computing (SciDAC) program under Award Number DE-AC02-05CH11231.  This research used resources of the National Energy Research Scientific Computing Center (NERSC), a U.S. Department of Energy Office of Science User Facility operated under Contract No. DE-AC02-05CH11231.
\end{acknowledgments}

%

\end{document}